\documentclass{ws-p8-50x6-00}
\def\be{\begin{equation}}
\def\ee{\end{equation}}
\def\ba{$\begin{array}{c}}
\def\ea{\end{array}$}
\def\o#1{{\cal O}(\lambda^#1)}

\begin{document}

\title{A new approach to texture analysis of quark mass matrices}

\author{Guo-Hong Wu}

\address{Institute of Theoretical Science, University of Oregon,
Eugene, OR  97403, USA\\E-mail: wu@dirac.uoregon.edu}


\maketitle

\abstracts{The triangular basis is proposed as an efficient 
way of analyzing general quark mass matrices. Applying the method
to hermitian hierarchical matrices with five texture zeros, analytic
predictions for quark mixing can be readily obtained. 
One unique texture pair is found most favorable with present
data. Some remarks are also made concerning parallel textures
between the up and down quark mass matrices with four zeros.}

\section{Mass matrices in the triangular basis}\label{sec:triangle}

Texture analysis of quark mass matrices\cite{FXreview}
 has been a subject of interest for over two decades, and
the advent of new data has rendered many popular textures
into  oblivion. A general and systematic approach based on the
observed mass and mixing spectrum becomes both welcome and necessary.
What we summarize here is one such approach using triangular matrices.

It is recently observed\cite{kmw1} that,
as a result of the chiral nature of the electroweak force
and the observed hierarchical structure 
among quark masses and mixing angles, the ten physical parameters
of the quark mass matrices
can be most simply encoded  in 
hierarchical matrices of upper triangular form. 
For example, in the basis where $M^U$ is diagonal,
$M^D$ is accurately given by,
\be \label{eq:TD}
M^U = \left(\begin{array}{ccc}
 m_u & & \cr
 & m_c & \cr
 & & m_t \end{array}\right) \;\;\;
M^D  = \left(\begin{array}{ccc}
m_d/V_{ud}^*&m_sV_{us}&m_bV_{ub}\cr
0&m_sV_{cs}&m_bV_{cb}\cr
0&0&m_bV_{tb} \end{array}\right) \cdot (1+\o4) \; ,
\ee
where $\lambda=|V_{us}|=0.22$.
Note that in this form, 
the unphysical right-handed rotations have been eliminated 
to a good approximation.

   Eq.(\ref{eq:TD}) is actually one of ten\cite{kmw2} triangular pairs
in the minimal-parameter basis (m.p.b.),
where each matrix element
consists of a simple product of a quark mass and certain
CKM elements, and the CP-violating weak-phase
is a linear combination of the phases of certain matrix elements. 
These ten pairs are simply related by weak basis transformation.
Given any quark mass matrices, one can read off their physical content
after converting them into one of the ten triangular pairs.
Alternatively, one can start from upper triangular matrices and
obtain mass matrices in other form, e.g. hermitian, and analyze
texture zeros in the new basis.

\section{Texture zeros of hermitian mass matrices}

   Based on the weak-scale quark mass relations 
$m_u:m_c:m_t \sim \lambda^8:\lambda^4:1$
and $m_d:m_s:m_b \sim \lambda^4:\lambda^2:1$, and on the hierarchical CKM
mixings
$V_{us} = \lambda$, $V_{cb} \sim \lambda^2$,
$V_{ub} \sim \lambda^4$, and $V_{td} \sim \lambda^3$,
we can write the
properly normalized Yukawa matrices
in the general hierarchical triangular form,
\begin{equation} \label{eq:triUD}
T^U = \left(\begin{array}{ccc}
       a_U \lambda^8 & b_U \lambda^6 & c_U \lambda^4 \\
          0          & d_U \lambda^4 & e_U \lambda^2 \\
          0          &    0          &  1
           \end{array}\right)\; ,
\;\;\;\;\;\;\;\;\;\;
T^D = \left(\begin{array}{ccc}
       a_D \lambda^4 & b_D \lambda^3 & c_D \lambda^3 \\
          0          & d_D \lambda^2 & e_D \lambda^2 \\
          0          &    0          &  1
           \end{array}\right)\; .
\end{equation}
Note that the diagonal elements are essentially the quark masses. 
The left-handed (LH) rotations are directly related to 
the off-diagonal elements,
whose coefficients are of order one
or much smaller to avoid fine-tuning\cite{PW} in getting the CKM
mixing.
This direct correspondence to masses and LH rotations is
a unique feature of upper triangular matrices, and
this feature makes the triangular matrices especially useful in 
analyzing quark mass matrices.
As an example, we now turn to the analysis of texture zeros
of hermitian quark mass matrices.

   Starting from the triangular matrices of Eq.~(\ref{eq:triUD}), we
can easily write down their corresponding hermitian form:
\begin{eqnarray} \label{eq:HU}
Y^U  & = & \left(\begin{array}{ccc}
       (a_U +  c_U c_U^* +
               b_U b_U^*/d_U ) \lambda^8 &
       (b_U + c_U e_U^*)  \lambda^6 & c_U \lambda^4 \\
       (b_U^* + c_U^* e_U)  \lambda^6  &
       (d_U + e_U e_U^*) \lambda^4 & e_U \lambda^2 \\
          c_U^* \lambda^4  & e_U^*\lambda^2   &  1
           \end{array}\right) \cdot (1+\o4)
\end{eqnarray}
\begin{eqnarray} \label{eq:HD}
Y^D & = & \left(\begin{array}{ccc}
      (a_D + b_D b_D^*/d_D) \lambda^4 &
       b_D \lambda^3 & c_D \lambda^3 \\
       b_D^* \lambda^3  & d_D \lambda^2 & e_D \lambda^2 \\
        c_D^* \lambda^3  &    e_D^* \lambda^2     &  1
           \end{array}\right) \cdot (1+\o2) \ .
\end{eqnarray}
It is seen that diagonal zeros in the hermitian matrices 
imply definite relations
between the diagonal and off-diagonal triangular parameters
(i.e. masses and LH rotation angles).
As a result of the different mass hierarchies in the up and 
down quark sector, there is a clear asymmetry 
between Eqs.~(\ref{eq:HU}) and (\ref{eq:HD}) regarding their texture zeros. 
For example, the $(2,2)$ element can be zero for $Y^U$ but not for $Y^D$;
both the $(1,1)$ and $(1,2)$ elements can vanish with $Y^U$ but not with 
$Y^D$.

   Hermitian mass matrix textures can then be analyzed as follows.
We first list all possible texture pairs ($M^U$, $M^D$) 
directly from Eqs.~(\ref{eq:HU}) and (\ref{eq:HD}).
Each pair is  then transformed into  one of the ten triangular
forms in the m.p.b. so that we can read-off
possible relations between quark masses and mixing.
Finally, the viability of each texture pair is tested by confronting
its predictions with data.
In this way, we find no viable hermitian pairs with six zeros,
including, {\em e.g.,} the Fritzsch texture\cite{fritzsch}.
 We are thus lead to consider hermitian mass matrices
with five texture zeros. 

\section{Hermitian mass matrices with 5 texture zeros}

  Following the procedure outlined above, we identify five candidates
for viable hermitian pairs with five texture zeros, as was first
obtained by Ramond, Roberts, and Ross (RRR) \cite{RRR}.
  We now  examine these five pairs analytically 
using  the m.p.b. triangular matrices, and confront 
their predictions with data. The running quark mass values
are taken from \cite{qmasses}, and we use a recent update\cite{exp} 
on $V_{ub}/V_{cb}$ and $V_{td}/V_{ts}$:
$\left| V_{ub}/V_{cb} \right|_{\rm exp} = 0.093 \pm 0.014$,
and $0.15 < \left| V_{td}/V_{ts} \right|_{\rm exp} < 0.24$. 
The results of this analysis~\cite{kmw2} can be summarized as follows.

   RRR patterns 1, 2, and 4 lead to the same predictions:
$\left|V_{td}/V_{ts} \right| \simeq \sqrt{m_d/m_s}=0.224 \pm 0.022$ and
$\left| V_{ub}/V_{cb} \right| \simeq \sqrt{m_u/m_c} = 0.059 \pm 0.006$.
Note that the numbers are independent of the scale at which the texture
is valid, and that
the latter prediction is on the low side and is disfavored by data.

 The 5th RRR pattern also gives rise to two relations, each with
two solutions depending on the sign of $d_U$. 
For the first relation,
\be \label{eq:V5us}
\left| V_{us} \right| \simeq
   \left| \sqrt{\frac{m_d}{m_s}} - e^{i\delta} \sqrt{\frac{m_u}{m_c}}
 \sqrt{\frac{V_{cb}^2}{(m_c/m_t) \pm V_{cb}^2}} \right| \ ,
\ee
where the relative phase is free as CP violation depends on additional 
phases in the mass matrices. Thus Eq.~(\ref{eq:V5us}) is valid
with a properly chosen phase.
For the second relation,
\be \label{eq:RRR5}
\left| \frac{V_{ub}}{V_{cb}} \right| \simeq 
\sqrt{\frac{m_u}{m_c} \left( \frac{m_c}{m_tV^2_{cb}} \pm 1 \right)}
= \left\{ \begin{array}{c} 0.107 \pm 0.012 \;\;\;\; (d_U>0, {\rm~~5a})
 \\ 0.068  \pm 0.011 \;\;\;\; (d_U<0, {\rm~~5b})
\end{array} \right .
\ee
where the numbers are given at $M_Z$ with $V_{cb}=0.040$.
If the texture is valid at a much higher scale like $M_{\rm GUT}$,
the weak scale predictions  for the quark mixing will decrease 
by only a few percent.
Comparing to data, we see that these predictions are
marginally acceptable.

 Finally, assuming the 3rd RRR texture to be valid at the weak
scale, two predictions result:
$\left| V_{ub} \right|  \simeq \sqrt{m_u/m_t} 
 = 0.0036 \pm 0.0004$,
and $\left| V_{us}/V_{cs} \right| \simeq \sqrt{m_d/m_s}
 = 0.224 \pm 0.022$.
Assuming the texture valid at the GUT scale will decrease the prediction
for $V_{ub}$ by a few percent.
  These predictions are in excellent agreement with data.

  In summary, the 3rd RRR five-texture-zero pattern is currently most
favorable.

\section{Remarks on parallel textures}

    There are some recent studies\cite{4zero}
 of 4-zero hermitian hierarchical mass matrices
with parallel textures between $M^U$ and $M^D$ (i.e. $M^U$ and $M^D$
have the same texture zeros in the same locations). 
In particular, the generalized Fritzsch texture with zeros at (1,1),
(1,3) and (3,1) has been a  focus of interest. 
Using triangular matrices, one can easily see that
generalized Fritzsch texture does not change the predictions
for $V_{ub}/V_{cb}$ and $V_{td}/V_{ts}$ from its 5-zero hermitian
counterparts (i.e. the 1st, 2nd, and 4th RRR patterns), and thus
not favorable with data. 
In fact, no parallel hermitian hierarchical
 textures with 4 zeros is found to be
favorable by current data\cite{ckw}.
In other words, the viable hermitian pairs display an asymmetry
between the up and down textures.
This asymmetry could serve as a useful guidline in building
realistic models for quark-lepton masses.

\section*{Acknowledgments}
 I would like to thank T.K. Kuo, S. Mansour,
and S.-H. Chiu for very enjoyable collaboration.


\begin{thebibliography}{99}

\bibitem{FXreview}
For a recent review, see
  H. Fritzsch and Z. Z. Xing, hep-ph/9912358, and references therein.


\bibitem{kmw1} T.K. Kuo, S. Mansour, and G.-H. Wu,
     \Journal{\PRD}{60}{093004}{1999}.

\bibitem{kmw2} T.K. Kuo, S. Mansour, and G.-H. Wu,
     \Journal{\PLB}{467}{116}{1999}.

\bibitem{PW}
R. D. Peccei and K. Wang, \Journal{\PRD}{53}{2712}{1996}.

\bibitem{fritzsch}
H.~Fritzsch,  \Journal{\NPB}{155}{189}{1979}.

\bibitem{RRR}
P.~Ramond, R.~Roberts, and G.~Ross, \Journal{\NPB}{406}{19}{1993}.

\bibitem{qmasses}
H.~Fusaoka and Y.~Koide, \Journal{\PRD}{57}{3986}{1998}.

\bibitem{exp}
  F. Parodi {\it et. al.,}  
{\em Nuovo Cim.} A {\bf 112}, 833 (1999);
   A. Ali and D. London, {\em Phys. Rep.} {\bf 320}, 79 (1999).

\bibitem{4zero}
H.~Fritzsch and Z.~Z.~Xing,\Journal{\NPB}{556}{49}{1999};
  M. Randhawa {\it et. al.,} \Journal{\PRD}{60}{051301}{1999};
  H. Nishiurai {\it et. al.,} \Journal{\PRD}{60}{013006}{1999};
 G. C. Branco {\it et. al.,} hep-ph/9911418.


\bibitem{ckw} 
  S.-H. Chiu, T.K. Kuo, and G.-H. Wu, hep-ph/0003224.


\end{thebibliography}
\end{document}